\documentclass[12pt,preprint]{aastex}
\begin{document}
\title{Constraining the IMF in Extreme Environments: \\
Detecting Young Low Mass Stars in Unresolved Starbursts}
\author{Michael R. Meyer and Julia Greissl}
\affil{Steward Observatory, The University of Arizona, Tucson, AZ \\ contact: mmeyer@as.arizona.edu}

\begin{abstract} 
We demonstrate the feasibility of detecting directly low mass
stars in unresolved super--star clusters with ages $<$ 10 Myr 
using near--infrared spectroscopy at modest resolution (R $\sim$ 1000). 
Such measurements could constrain the ratio of high to low mass stars
in these extreme star--forming events, providing a direct test on 
the universal nature of the initial mass function (IMF) compared to the
disk of the Milky Way (Chabrier, 2003).  We compute
the integrated light of super--star clusters with masses of 
10$^6$ M$_{\odot}$ drawn from the Salpeter
(1955) and Chabrier (2003) IMFs for clusters aged 1, 3, and 10 Myr.
We combine, for the first time, results from Starburst99 
(Leitherer et al. 1999) for the main sequence and post--main 
sequence population (including nebular emission) 
with pre--main sequence (PMS) evolutionary models (Siess et al. 2000) 
for the low mass stars as a function of age.  
We show that $\sim$ 4--12 \% of the integrated light observed
at 2.2 $\mu$m comes from low mass PMS stars with late--type stellar
absorption features at ages $<$ 3 Myr.  This light is discernable 
using high signal--to--noise spectra ($>$ 100) at R=1000 placing
constraints on the ratio of high to low mass stars contributing
to the integrated light of the cluster.  
\end{abstract}


\section {Introduction}

The observed initial mass function of stars, averaged over space
and time in the disk of the Milky Way \footnote{See for example 
the review of Chabrier (2003) and references therein.}, provides
a fundamental test for any theory of star formation.  Characteristic 
masses (such as the transition from a power--law IMF to a log--normal 
form, as well as a mean mass) provide evidence for preferred scales
of star formation (Adams and Fatuzzo, 1996; Larson, 1985; 2005).  The stellar 
IMF observed in nearby ($<$ 1 kpc) star--forming regions appears
to be broadly consistent with having been drawn from the field star
IMF (Meyer et al. 2000).  As a result, astronomers have been forced
to look for differences either at very low masses (e.g. Briceno
et al. 2002) or more extreme galactic star--forming environments 
(e.g. Figer et al. 1999).  

Studies of starburst galaxies, with star 
formation rates orders of magnitude higher than normal disk galaxies 
(e.g. Kennicutt, 1998), have led to claims of 
IMFs distinct from that characterizing the disk of the Milky 
Way.  For example, Smith and Gallagher (2001) find that some 
massive clusters in M82 exhibit a ``low mass cut--off''. 
However, F{\"o}rster Schreiber et al. (2003; see also McCrady et al. 2005; 
Rieke et al.  1993) find a power--law
slope for stars $>$ 10 M$_{\odot}$ with a flattening at masses $<$ 
1.0 M$_{\odot}$ similar to the IMF of Chabrier (2003; see also Kroupa, 
2001).  Recent studies of the integrated light of super--star 
clusters forming in interacting galaxies suggest variations in 
mass--to--light ratios that could be interpreted as variations
in the IMF (Mengel et al. 2002).  Since such systems might be 
analogues for the vigorous star formation thought to have occurred
in the early universe (e.g. Madau et al. 1996) it is particularly 
important to understand whether the IMF varies with metalicity or 
star--formation rate.

In order to: 1) provide important constraints on theories of 
star formation under extreme conditions; and 2) test claims of 
variations in the IMF for starburst galaxies and super--star 
clusters, we explore the feasibility of detecting directly 
the low mass pre--main sequence stars in young stellar populations. 
In section II, we outline our approach to simulating the integrated
near--IR light from very young rich star--forming events.
In section III, we discuss our results as a function of input model 
assumptions, and in section IV we discuss the uncertainties in 
constraining the ratio of high to low mass stars using the technique
described here. 

\section{The Approach} 

We have simulated the integrated near--infrared
spectrum from a stellar population using the assumptions
outlined below.  First, we assume single star--forming events 
with total mass 10$^6$ M$_{\odot}$ populated in Monte Carlo fashion 
by either a Salpeter (1955; hereafter S55) or Chabrier (2003; 
hereafter C03) IMF from 0.1--100 
M$_{\odot}$ at 
ages of 1, 3, and 10 Myr.  We take into account the effects of 
main sequence and post--main sequence evolution using Starburst99
(Leitherer et al. 1999).  This program also includes nebular
emission that results from interaction of the ionizing flux from 
the most massive stars formed with the surrounding medium as it
evolves along with the ensemble radiation field.  Because we 
are interested in detecting low mass stars directly as a fraction
of the integrated light in the near--infrared, 
the high mass stars and nebular emission
\footnote{Dominated by free--free emission with a characteristic spectrum 
of $F_{\lambda} \sim \lambda^{-2}$.} simply 
provide a featureless continuum against which we wish to detect
spectral features from late--type stars.  

We also include, 
for the first time, the effects of {\it pre--main sequence} (PMS) evolution
explicitly on the integrated light of the stellar population.  
We use the models of Seiss et al. (2000) to provide appropriate
mass--luminosity relationships as a function of time over a broad
range of stellar masses.  For populations
1/3/10 Myr old, all stars $<$ 7/5/3 M$_{\odot}$ are PMS respectively. 
This corresponds to 0.971/0.956/0.911 \% of all stars with total masses of 
of approximately 0.69/0.63/0.51 $\times$ the cluster mass 
in the PMS component for 
1/3/10 Myr respectively (assuming a Chabrier, 2003 IMF). 
We adopt tables of intrinsic colors and bolometric corrections
as a function of effective temperature and spectral type 
from Cox (2000) and use
these to estimate the contribution of each PMS star to the integrated
J-- (1.25 $\mu$m), H-- (1.65 $\mu$m), and K--band (2.2 $\mu$m) light. 
For each star, a mass is assigned based on probabilities from the 
input IMF. 
For the assumed age, this translates into a temperature and bolometric
luminosity.  For the model temperature, a color and bolometric 
correction is adopted resulting in a calculated absolute M$_J$, 
M$_H$, and M$_K$ magnitude.   Based on this temperature, we select 
a continuum normalized SNR $\sim$ 50 spectrum at R=1000 from the library of Meyer (1996) 
with weighting 
according to its absolute flux to produce the integrated PMS spectrum
for the low mass stars.  We then combined the PMS spectrum
with the appropriately flux--weighted continuum from Starburst99
for the high mass main sequence and post--main sequence stars, 
along with the nebular continuum to produce the total integrated
spectrum of the population as a function of age (Figure 1). 

\section{The Results}

The integrated PMS spectra shown in Figure 1 exhibit 
late--type features due to CaI at 2.26 $\mu$m and the first--overtone CO absorption
at 2.29 and 2.32 $\mu$m (prominent in stars $>$ K0) as well 
as features of MgI at 2.28 $\mu$m (seen in stars from F2--M1) 
as noted by Ali et al. (1995), 
Hinkle and Wallace (1997), and Kleinmann and Hall (1986).  While 
these features are difficult to see against
the continuum of the high mass stars and nebular
emission, they are detected at the 
1 \% level in high signal--to--noise
ratio spectra.  Table 1 shows the breakdown of relative flux
in the J--, H--, and K--bands as a function of age and input IMF. 
The nebular continuum dominates at the youngest ages and becomes
less important with time (Leitherer et al. 1995).  
However, the low mass stars are also 
brightest at the youngest ages due to their PMS nature, 
partially off--setting the effect
of strong nebular continuum.  It appears that the low mass 
PMS stars contribute a roughly constant fraction to the integrated
K--band flux from $<$ 1 to $>$ 3 Myr ($>$ 0.07 to $<$ 0.04 for 
the Chabrier (2003) IMF) though this effect is commonly ignored.
Once the most massive stars 
evolve to become M supergiants at an age of $\sim$ 8 Myr years, they
dominate the continuum and the PMS stars become $<$ 1 \% of the 
integrated K--band flux making them almost impossible to detect.
However, at ages $<$ 8 Myr, the low mass stars are detectable
through the CO absorption features against the continuum of the 
high mass stars and nebular continuum.  

We have investigated the diagnostic power of the spectral features
indicated in Figure 1 by measuring the equivalent width of 
the features as a function of age and input IMF using the SPLOT
routine within IRAF.  Variations
in determining the local continuum for each spectral feature 
provide the error estimates in the equivalent widths. 
As can be seen 
in Table 2, the combined strength of the (CaI+CO) indices, which 
are dominated by stars cooler than 3500 K and less than 0.5 M$_{\odot}$ 
from ages 1--3 Myr (Siess et al. 2000), decrease
by factors of 2.13 (S55) and 3.28 (C03) as the populations age 
and the low mass stars become less luminous.  Note that the combined index
evolves more from 1--3 Myr for C03 (shallower IMF) compared to S55 
since it is more sensitive to stars 0.2--0.5 M$_{\odot}$, whose
luminosities are evolving faster than the stars $<$ 0.2 M$_{\odot}$. 
The MgI index, tracing
stars with temperatures between 3700--7000 K, is the same
for both IMFs at 1 Myr where it is dominated by stars 
$>$ 1.0 M$_{\odot}$ (Siess et al. 2000): at high masses the IMF
slopes are the same. 
Again because this index is dominated by higher mass stars from the
C03 IMF whose luminosities are changing faster, it changes more 
from 1--3 Myr compared to S55 (factors of 3.36 versus 2.61).  
Note that the mass range sampled by the MgI feature, attributed to 
warmer stars compared to the CaI+CO index, decreases with time
from with the upper mass limit going from 4 M$_{\odot}$ (1 Myr) 
to 2.5 M$_{\odot}$ (3 Myr).  While the equivalent width ratio 
given in the last column of Table 2
$(EW[CaI]+EW[CO 2-0)])/EW[MgI]$ does not evolve significantly
from 1--3 Myr for the Chabrier (2003) IMF, both values are
distinguished from the range of values realized from the 
S55 IMF.  The difference between the 1 Myr S55 value
and the 3 Myr old C03 value is $8.98 \pm 0.82 - 6.44 \pm 0.30 = 
2.54 \pm 0.87$.  Although the SNR of the spectra required
in order to accurately measure the weak MgI feature probably
exceeds 300 at R = 1000, our simulations suggest that we
can distinguish between the C03 and S55 
IMFs at the $\sim$ 3 $\sigma$ level provided we can estimate the age
of the stellar population to within a factor of three. 

\section{The Discussion}

We have demonstrated that, given our model assumptions are
correct, it should be possible to discern the presence of the low mass 
stars in a very young unresolved stellar population.  This would
place important constraints on the ratio of high to low mass stars, 
and thus the IMF, in extreme star--forming environments. 
We now examine how sensitive the conclusions presented above 
are to the assumptions we have made.   
Would utilizing different 
PMS evolutionary calculations make a difference? 
Replacing the mass--luminosity--temperature relationships 
of Siess et al. (2000) as a function of age with those of Palla 
and Stahler (2000) or D'Antona Mazzitelli (1994) 
would result in a change of $<$ 10 \% in the absolute 
K--band magnitude of a 1 M$_{\odot}$ star at an age of 1 Myr.
While there are larger uncertainties in the PMS evolutionary models
at lower masses (e.g. Hillenbrand and White, 2004), the qualitative 
behavior of the models within our simulations will be the same. 

We have assumed that Starburst99 correctly predicts the amount of
nebular continuum for the unresolved population.  The prescription
used is described in Leitherer et al. (1995) and assumes that
all ionizing photons contribute with high efficiency to producing
the observed free--free emission.  
We considered the effects of changing the maximum stellar
mass from 100 M$_{\odot}$ to 130 and 200 M$_{\odot}$ in our model stellar 
clusters.  Moving the maximum mass to 130 M$_{\odot}$ made only a 
1.3 \% decrease to our prediction for the integrated K--band contribution 
from PMS stars to 5.7 \% (increasing the continuum dilution of the late--type 
absorption features), while increasing the maximum mass to 200 M$_{\odot}$ 
decreased the integrated K--band light to 5.4 \% for a 1 Myr old population
assuming a Chabrier (2003) IMF (cf. Table I).  Oey and Clarke (2005) 
have presented evidence for a universal maximum stellar mass between 120--200 
M$_{\odot}$ while Figer (2005) suggests a maximum 
observed stellar mass of 130 M$_{\odot}$ based on observations of
the Arches cluster (see Massey, 1999 for an alternative point of view). 
Further, if, as is observed in some
star--forming complexes, the HII region has broken out of the 
molecular cloud over an appreciable solid angle, the estimated
free--free continuum is an upper limit, making our predicted
line strengths for low mass stars a lower limit. 

We have also neglected the presence of circumstellar disks which 
could produce an additional veiling continuum further diluting 
the observed late--type stellar features from low mass stars.
However, not all young stellar objects exhibit excess continuum 
emission at 2.3 $\mu$m and typical excess emission appears to 
peak at longer wavelengths (Meyer et al. 1997; Muzerolle
et al. 2003).  Further we have also underestimated the absorption
line strength of the CO feature, which is surface gravity sensitive, 
as we have used higher gravity dwarf star templates to represent
lower gravity PMS objects.  These two effects tend to cancel. 
How does our choice of R=1000 affect the detectable of the low 
mass stars?  While lower spectral resolution might make it 
easier to detect the spectral features 
(Nyquist sampled at R $>$ 300) at the required SNR, it may be important
to obtain higher resolution spectra in order to constrain 
the surface gravity of the dominant stellar population
responsible for the stellar absorption.  If any 
supergiant stars contribute to the integrated light
observed at 2.3 $\mu$m, R $<$ 3000 spectra should enable
their identification due to the extremely low surface gravity, 
to which CO is sensitive (Aaronson et al. 1978; Meyer et al. 1998). 
In this preliminary work, we have ignored the effects of metalicity.
Starburst galaxies are thought to have higher than solar 
metalicity, which would increase the absorption line strengths
predicted in our simulations. 

One degeneracy in constraining the slope of the IMF with 
which we must contend is that between the ratio of high
to low mass stars and the luminosity of the PMS population
as a function of age (see Table 1).  If one assumed an 
age of 3 Myr for what was actually a 1 Myr old cluster, 
one might erroneously favor a steeper IMF (e.g. S55) 
rather than a C03 IMF based on analysis of the Ca+CO 
absorption strength alone.  However, the
line ratio analysis outlined above can distinguish 
between IMFs with powerlaw indices that differ by $\sim$ 
1.0 dex above and below 1 M$_{\odot}$ (e.g. between S55 and C03)
provided that the age of the stellar population can be
constrained within a factor of $\times$ 3.  
Recent work on modelling the spectral energy distributions (SEDs) of 
starburst galaxies suggest that ages for them can be derived within 
factors of $\times$ 3 given $>$ four points on the SED
including the U-- and B--bands as well as near--IR data 
(Anders et al. 2004).  In addition, 
it is vital to know whether or not the cluster might contain
supergiants (corresponding to ages $>$ 8 Myr) which will 
dominate the near--IR continuum once they appear. 
Vazquez and Leitherer (2005) present model SEDs
for very young starbursts that indicate it is relatively
straight--forward to distinguish clusters $<$ and $>$ 10 Myr
based on the energy emitted between 300--1000 $\AA$ despite
the uncertainties in the theoretical models.  Further, Gon\'zalez Delgado
and P\'erez (2000) show that results from SED modelling
which include UV data agree well with other techniques and
give a consistent age of 3 Myr for the stellar populations
associated with the HII region NGC 604 in M 33.  

How can we treat uncertainties in the level of free--free 
emission?  Rather than rely on the model predictions outlined above 
for the relative ratios of nebular emission to stellar
emission, it would be very valuable to constrain the
amount of nebular emission through observations.  Since
free--free emission from such a young stellar population
becomes increasingly dominant at longer wavelengths, one
could measure the free--free emission directly at centimeter
wavelengths where it is expected to dominate dust and 
compare the inferred amount in the near--IR to the predictions 
of Starburst99.  Spatially resolved observations
of the HII regions in M17 by Ando et al. (2002) indicate 
qualitative agreement between observed IR emission and 
predicted free--free emission extrapolated from centimeter
observations.  
An additional check on this extrapolation would be the
predicted reddening corrected Br $\gamma$ line flux given 
the observed free--free emission.    
If one is able to detect the weak MgI feature discussed above, 
the line ratio suggested in Table 2 would render the measurement
insensitive to uncertainties in the continuum veiling due to 
free--free emission.  Removal
of the free--free contribution to the observed near--IR spectrum
and a model of the integrated stellar spectrum would place
a direct constraint of the ratio of high to low mass stars.

Such constraints are sensitive to the 
relative slope of the IMF between 10--100 M$_{\odot}$ and 
0.1--1.0 M$_{\odot}$, crucial for precise estimates of the 
mass--to--light ratios for older stellar populations.  Is this
observational approach feasible?  Several candidate super--star clusters
found in the overlap region of the interacting galaxy pair NGC 4038/39 
exhibit weak CO absorption and are thought to be $<$ 10 Myr old
(Mengel et al. 2002) indicating that they lack late--type supergiants. 
These targets have K--band magnitudes between 
14$^m < K < 16^m$ (Kassin et al.  2003).  We 
estimate that it would take $\sim$ 10 hours at R $\sim$ 1000 to obtain
SNR $\sim$ 300 in the K--band to detect late--type features
from low mass stars in a $K \sim 14^m$ young super--star cluster
on an 8 meter telescope.  
Added information available through 
spectral observations in the J-- and H--bands may make application of
this technique easier.  
Despite all the caveats listed above, we believe the experiment 
is worthwhile, given the important implications of the observations. 

We would like to thank the Claus Leitherer for the helpful suggestion
that we investigate the effects of free--free emission on the 
integrated spectra of starbursts at young ages, as well as an 
anonymous referee for suggestions that improved the presentation
of this paper.  We also thank the 
organizers of the IMF@50 conference (Evidge Corbelli, Franscesco
Palla, and Hans Zinnecker) for the opportunity to present a 
preliminary version of this work at the Abetto de Spinoza, 
and Morten Andersen along with Kelsey Johnson 
for comments on a draft of the paper. 
We gratefully acknowledge the support of a Cottrell Scholar
award from the Research Corporation as well as NASA grant 
GO--9846 from the Space Telescope Science Institute.

\begin{table}
\begin{small}
\centering
\caption{Fraction of Integrated PMS Light for 10$^6$ M${\odot}$ Cluster} 
\begin{tabular}{lllllc}
\multicolumn{1}{c}{Band}   &
\multicolumn{1}{c}{$IMF^1$} &
\multicolumn{1}{c}{$\tau$ (age) } &
\multicolumn{1}{c}{$F_{PMS}^2$}&
\multicolumn{1}{c}{$F_{MS}$}&
\multicolumn{1}{c}{$F_{NEB}$} \\\hline\hline
J & S55 & 1 Myr & 0.13 & 0.17 & 0.70  \\
H & S55 & 1 Myr & 0.175 & 0.125 & 0.70  \\
K & S55 & 1 Myr & 0.12  & 0.07 & 0.81  \\   
J & C03 & 1 Myr & 0.09 & 0.18 & 0.73  \\
H & C03 & 1 Myr & 0.10 & 0.14 & 0.76  \\
K & C03 & 1 Myr & 0.07 & 0.07 & 0.86  \\   
J & C03 & 3 Myr & 0.04 & 0.44 & 0.52  \\
H & C03 & 3 Myr & 0.05 & 0.37 & 0.58  \\
K & C03 & 3 Myr & 0.04 & 0.23 & 0.73  \\   
J & C03 & 10 Myr & 0.01 & 0.99 & 0.00  \\
H & C03 & 10 Myr & 0.01 & 0.99 & 0.00  \\
K & C03 & 10 Myr & 0.01 & 0.99 & 0.00  \\   
\hline 
\multicolumn{6}{l}{\footnotesize $^1$
S55 indicates Salpeter (1955) and C03 indicates Chabrier (2003) IMF.}\\
\multicolumn{6}{l}{\footnotesize $^2$
Pre--main sequence mass--luminosity taken from Seiss et al. (2000).} \\
\label{light}
\end{tabular}
\end{small}
\end{table}

\begin{table}
\begin{small}
\centering
\caption{Equivalent Widths of Late--Type Features Visible in Integrated Spectra} 
\begin{tabular}{lllllc}
\multicolumn{1}{c}{$IMF^1$} &
\multicolumn{1}{c}{$\tau$ (age) } &
\multicolumn{1}{c}{$EW[CaI]^2$}&
\multicolumn{1}{c}{$EW[MgI]^2$}&
\multicolumn{1}{c}{$EW[CO(2-0)]^2$} & 
\multicolumn{1}{c}{$(EW[CO(2-0)] + EW[CaI])/EW[MgI]$} \\\hline\hline
 S55 & 1 Myr & 0.3168 $\pm$ 0.0188 & 0.1092 $\pm$ 0.0090 & 0.6632 $\pm$ 0.0322 &8.98 $\pm$ 0.82 \\
 S55 & 3 Myr & 0.1515 $\pm$ 0.0083 & 0.0418 $\pm$ 0.0011 & 0.3090 $\pm$ 0.0114 &11.01 $\pm$ 0.44 \\
 C03 & 1 Myr & 0.2055 $\pm$ 0.0067 & 0.1150 $\pm$ 0.0079 & 0.5162 $\pm$ 0.0103 &6.27 $\pm$ 0.44 \\
 C03 & 3 Myr & 0.0640 $\pm$ 0.0016 & 0.0342 $\pm$ 0.0012 & 0.1558 $\pm$ 0.0062 &6.44 $\pm$ 0.30 \\
\hline 
\multicolumn{6}{l}{\footnotesize $^1$
S55 indicates Salpeter (1955) and C03 indicates Chabrier (2003) IMF.}\\
\multicolumn{6}{l}{\footnotesize $^2$
Equivalent widths in microns measured as described in the text.} \\
\label{ew}
\end{tabular}
\end{small}
\end{table}

\begin{figure}
\caption{Integrated spectra for a 10$^6$ M$_{\odot}$ 
cluster distributed from 0.1--100 M$_{\odot}$ via the 
S55 IMF for ages of 1 and 3 Myr compared to 
similar spectra derived from the C03 IMF at comparable ages 
based on the mass-luminosity relationships
of Seiss et al. (2000).  Late--type features due to the presence
of low mass pre--main sequence stars, visible at the 1 \% level
against the light of the higher mass main sequence stars and
nebular continuum, are noted. The presence of residual telluric absorption
at 2.3150 $\mu$m is also indicated.}
\label{spectra}
\end{figure}

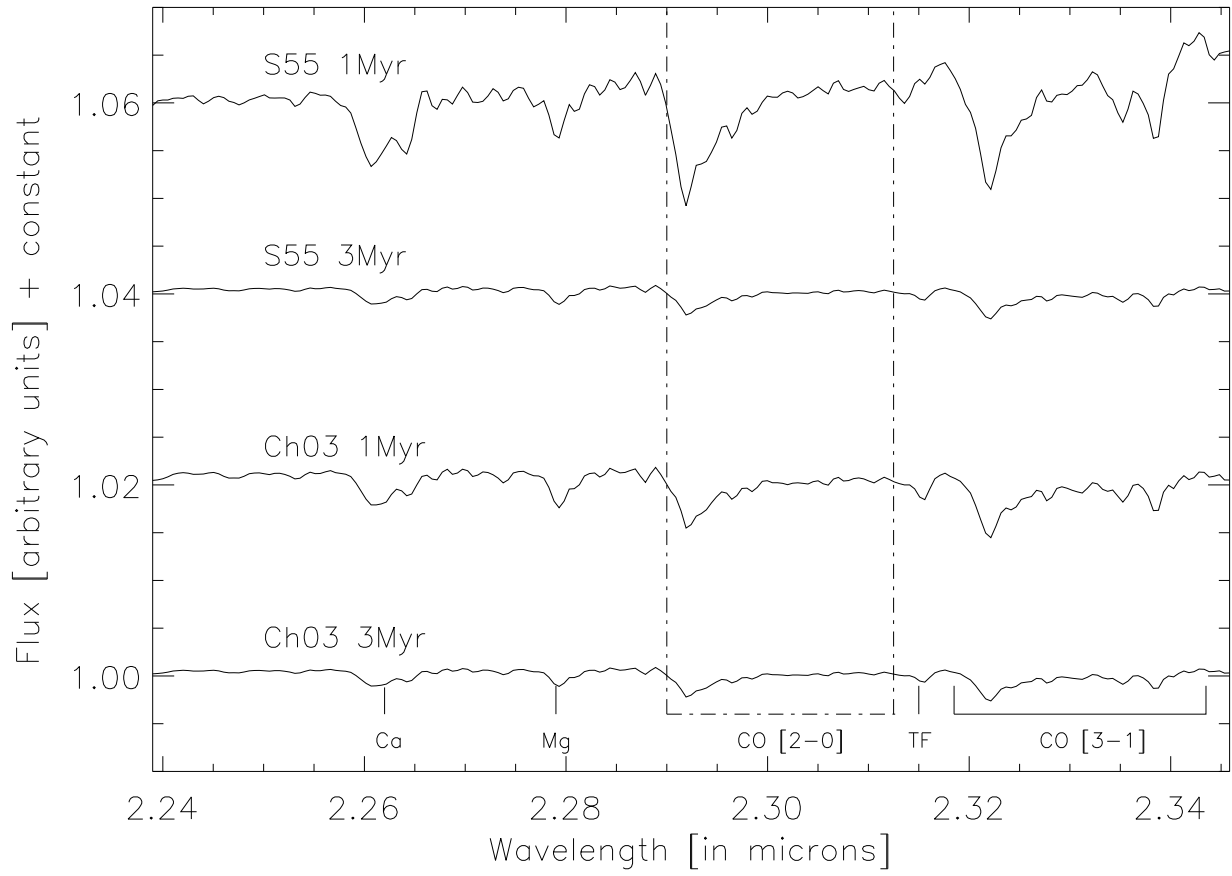
\end{document}